\documentclass{article}
\usepackage{amssymb}
\usepackage{amsmath}
\usepackage[dvips]{graphicx}

\title{The Large scale instability in rotating fluid with small scale force}

\author{$^{1,2}$\textbf{M.I.Kopp}, $^3$\textbf{A.V.Tur}, $^{1,2}$\textbf{V.V.Yanovsky}}

\date{}

\begin{document}

\maketitle

$^{1}$ \textit{Institute for Single Crystals, National Academy of
Science Ukraine, Kharkov 61001, Ukraine}

$^{2}$\textit{V.N.Karazin Kharkiv National University 4 Slobody Sq., Kharkov 61022, Ukraine}

$^{3}$\textit{Universit\'{e} de Toulouse [UPS], CNRS, Institut de Recherche en Astrophysique et Plan\'{e}tologie,
9 avenue du Colonel Roche, BP 44346, 31028 Toulouse Cedex 4, France}

\abstract{In this paper, we find a new large scale instability displayed by a rotating flow in forced turbulence.
The turbulence is generated by a small scale external force at low Reynolds number. The theory is built on the rigorous asymptotic method of multi-scale development. The nonlinear equations for the instability are obtained at the third order of the perturbation theory. In this article, we explain a detailed study of the nonlinear stage of the instability and generation vortex kinks.}

\bigskip


\bigskip

\textbf{Pacs}: 47.32C-; 47.27.De;47.27.em; 47.55.Hd.

\section{Introduction}
\label{i1}

It is well known, that the rotating effects play an important role in many practical and theoretical applications for fluid mechanics  \cite{s1} and are especially important for geophysics and astrophysics \cite{s2}, \cite{s3}, when one have to deal with rotating  objects such as Earth, Jupiter, Sun etc. Rotating fluids could generate different wave and vortex motions. For example, gyroscopic waves, Rossby waves, internal waves, located vortices and coherent vortex structures \cite{s4}-\cite{s7}.  Among the vortex structures the most interesting are the large scale ones, since they carry out the efficient transfer of energy and impulse. The structures which have characteristic scale much more than the scale of turbulence or of the external force which generates this turbulence are understood as large scale ones. At present we can note, that there are a lot of instabilities which generate the large scale vortex structures (see for example \cite{s8}-\cite{s14}), including  in rotating fluid with the non-homogeneous turbulence  \cite{s15}. In this work we find new large scale instability in rotating fluid, under the impact of small external force which keeps up turbulent fluctuations.  The nonlinear big scale helical vortex structures of Beltrami type or localized kinks with internal helical structure appear as a result of the development of this instability in rotating fluid.  We can consider that external small scale force substitutes the action of small scale turbulence.  It is supposed that external force is in plane $(X, Y)$, which is perpendicular to the rotation axis, i.e. axis $Z$ is directed along the vector of angular velocity of rotation $\vec{\Omega }$. Helical $2D$ field of velocity  $W_{x}$, $W_{y} $ turns around axis $Z$, when $Z$ changes in the kink which links the hyperbolic point and the stable focus (fig.\ref{f2}).  Moreover this field does some turns in the kink, which links instable and stable focuses (Fig.\ref{f3}). The found instability belongs to the class of instabilities called hydrodynamic $\alpha $-effect. For these instabilities the positive feedback between velocity components of following type is typical:
\[\partial _{T} W_{x} -\Delta W_{x} -\alpha _{y} \frac{\partial }{\partial z} W_{y} =0,\]

\[{\partial _{T} W_{y} -\Delta W_{y} +\alpha _{x} \frac{\partial }{\partial z} W_{x} =0,}\]
what leads to the instability. $\alpha$-effect origins from  magnetic hydrodynamics, where it engenders the increase of large scale magnetic fields (see for example  \cite{s16}). Afterwards it was generalized on usual hydrodynamics. Some examples of hydrodynamics $\alpha $-effect \cite{s8}-\cite{s14} known for the time being. From this point of view, a new example of $\alpha $-effect is found in this work. The theory of this instability is developed rigourously using the method of asymptotic multi-scale development similar to what was done by Frisch, She and Sulem for the theory of the AKA effect \cite{s13}. This method allows to find the equations for large scale perturbations as the secular equations of perturbation theory, to calculate the Reynolds stress tensor and to find the instability. The small parameter of asymptotical development is the number of Reynolds $R$, $R\ll 1$. Our paper is organised as follows: in Section \ref{i2} we formulate the problem and the main equations in rotating system of coordinates; in Section \ref{i3} we examine the principal scheme of the multi-scale development and we give the secular equations. In Section \ref{i4} we calculate the velocity field of zero approximation.  In Section \ref{i5} we describe the calculations of the Reynolds stress and find the large scale instability. In Section \ref{i6} we discuss the saturation of the instability and find non linear stationar vortex structures. The results obtained are discussed in the conclusions given in Section \ref{i7}.

\section{The Main Equations and Formulation of the Problem}
\label{i2}

Let us examine the equations of motion for non-compressible rotating fluid with external force  $\vec{F}_{0} $ in rotating coordinates system :
\begin{equation}\label{EQ1}
    \frac{\partial \vec{V}}{\partial t} +\left(\vec{V}\nabla \right)\vec{V}+2\vec{\Omega }\times \vec{V}=-\frac{1}{\rho _{0} } \nabla P+\nu \Delta \vec{V}+\vec{F}_{0} ,
\end{equation}

\begin{equation}\label{EQ2}
    div\vec{V}=0
\end{equation}
The external force $\vec{F}_{0}$ is divergence-free. Here $\vec{\Omega}$-angular velocity of fluid rotation, $\nu$-viscosity, $\rho _{0}$-constant fluid density. Let us design characteristic amplitude of force $f_0$, and its characteristic space and time scale $\lambda _{0}$ and $t_{0}$ respectively.

Then $\vec{F}_{0} = f_0 \vec{F}_{0} $$\left(\frac{\vec{x}}{\lambda _{0} }, \frac{t}{t_{0} } \right)$.  We will design the characteristic amplitude of velocity, generated by external force as $v_0$. We choose the dimensionless variables $(t, \vec{x}, \vec{V})$:

\[\vec{x}\to \frac{\vec{x}}{\lambda _{0} } ,\quad t\to \frac{t}{t_{0} } ,\quad \vec{V}\to \frac{\vec{V}}{v_{0} } ,\quad \vec{F}_{0} \to \frac{\vec{F}_{0} }{f_{0} } ,\quad P\to \frac{P}{\rho _{0} P_{0} } ,\]

\[t_{0} =\frac{\lambda _{0}^{2} }{\nu } ,\quad P_{0} =\frac{v_{0} \nu }{\lambda _{0} } ,\quad f_{0} =\frac{v_{0} \nu }{\lambda _{0}^{2} } ,\quad v_{0} =\frac{f_{0} \lambda _{0}^{2} }{\nu } .\]
Then, in dimensionless variables the equation \eqref{EQ1} takes forme~:

\begin{equation} \label{EQ3}
  \frac{\partial \vec{V}}{\partial t} +R(\vec{V}\cdot \nabla )\vec{V}+\vec{D}\times \vec{V}=-\nabla P+\Delta \vec{V}+\vec{F}_{0} ,
\end{equation}
$\vec{D}=2 \frac{\vec{\Omega} \lambda^2_0}{\nu}$, $R=\frac{\lambda _{0} v_{0} }{\nu}$, $\left|\vec{D}\right|=\sqrt{Ta}$ where $R$ and $Ta=\frac{4\Omega ^{2} \lambda _{0}^{4} }{\nu ^{2}}$ are respectively the Reynolds number and the Taylor number on scale $\lambda _{0}$. Further we will consider the Reynolds number as small $R \ll 1$ and will construct on this small parameter the asymptotical development. Concerning the parameter $D$, we do not choose any range of values for the moment. Let us examine the following formulation of the problem. We consider the external force as being small scale  and of high frequency. This force leads to small scale fluctuations in velocity. After averaging, these quickly oscillating fluctuations vanish. Nevertheless, due to small nonlinear interactions in some orders of perturbation theory, nonzero terms can occur after averaging. This means that they are not oscillatory, that is to say, they are large scale. From a formal point of view, these terms are secular, i.e., they create the conditions for the solvability of a large scale asymptotic development. So the purpose of this paper is to find and study the solvability equations, i.e., the equations for large scale perturbations. Let us denote the small scale variables by $x_{0} =(\vec{x}_{0}, t_{0})$ , and the large scale ones by $X=(\vec{X},T)$. The small scale partial derivative operation $\frac{\partial }{\partial x_{0}^{i}}$, $\frac{\partial }{\partial t_{0}}$, and the large scale ones $\frac{\partial }{\partial \vec{X}}$, $ \frac{\partial }{\partial T}$ are written, respectively, as $\partial _{i}$, $\partial _{t}$, $\nabla _{i}$ and $\partial _{T}$. To construct a multi-scale asymptotic development we follow the method which is proposed in \cite{s16}.

\section{The multi-scale asymptotic development}
\label{i3}

Let us search for the solution to equations \eqref{EQ2} and \eqref{EQ3} in the following form:

\begin{equation} \label{EQ4}
  \vec{V}(\vec{x},t)=\frac{1}{R} \vec{W}_{-1} (X)+\vec{v}_{0} (x_{0} )+R\vec{v}_{1} +R^{2} \vec{v}_{2} +R^{3} \vec{v}_{3} +\cdots ,
\end{equation}

\begin{equation} \label{EQ5}
  T(\vec{x},t)=\frac{1}{R} T_{-1} (X)+T_{0} (x_{0} )+RT_{1} +R^{2} T_{2} +R^{3} T_{3} +\cdots ,
\end{equation}
\[P(\vec{x},t)=\frac{1}{R^{3} } P_{-3} (X)+\frac{1}{R^{2} } P_{-2} (X)+\frac{1}{R} P_{-1} (X)+P_{0} (x_{0} )+\]
\begin{equation} \label{EQ6}
   +R(P_{1} +\overline{P}_{1} (X))+R^{2} P_{2} +R^{3} P_{3} +\cdots .
\end{equation}
Let us introduce the following equalities: $\vec{X}=R^{2} \vec{x}_{0}$ and $T=R^{4} t_{0}$ which lead to the expression for the space and time derivatives:

\begin{equation} \label{EQ7}
   \frac{\partial }{\partial x^{i} } =\partial _{i} +R^{2} \nabla _{i} ,
\end{equation}

\begin{equation} \label{EQ8}
   \frac{\partial }{\partial t} =\partial _{t} +R^{4} \partial _{T} ,
\end{equation}

\begin{equation} \label{EQ9}
   \frac{\partial ^{2} }{\partial x^{j} \partial x^{j} } =\partial _{jj} +2R^{2} \partial _{j} \nabla _{j} +R^{4} \partial _{jj} .
\end{equation}
Using indicial notation, the system of equation can be written as
\[{(\partial _{t} +R^{4} \partial _{T} )V^{i} +R(\partial _{j} +R^{2} \nabla _{j} )(V^{i} V^{j} )+D^{j} \varepsilon _{ijk} V^{k} =}\]
\begin{equation}\label{EQ10}
    =(\partial _{j} +R^{2} \nabla _{j} )P+(\partial _{jj} +2R^{2} \partial _{j} \nabla _{j} +R^{4} \nabla _{jj} )V^{i} +F_{0}^{i} ,
\end{equation}

\begin{equation} \label{EQ11}
   \partial _{t} T-\partial _{jj} T=-V^{z} -R\partial _{j} \left(V^{i} T\right),
\end{equation}

\begin{equation} \label{EQ12}
  \left(\partial _{i} +R^{2} \nabla _{i} \right)V^{i} =0.
\end{equation}
Substituting these expressions into the initial equations \eqref{EQ2} and \eqref{EQ3} and then gathering together the terms of the same order, we obtain the equations of the multi-scale asymptotic development and write down the obtained equations up to order $R^{3}$ inclusive. In the order $R^{-3}$ there is only the equation

\begin{equation} \label{EQ13}
  \partial _{i} P_{-3} =0,\Rightarrow P_{-3} =P_{-3} (X).
\end{equation}
In order $R^{-2}$ we have the equation

\begin{equation} \label{EQ14}
  \partial _{i} P_{-2} =0,\Rightarrow P_{-2} =P_{-2} (X).
\end{equation}
In order $R^{-1} $ we get a system of equations:

\begin{equation} \label{EQ15}
   \partial _{t} W_{-1}^{i} -\partial _{jj} W_{-1}^{i} +D^{j} \varepsilon _{ijk} W_{-1}^{k} =-(\partial _{i} P_{-1} +\nabla _{i} P_{-3} )-\partial _{j} W_{-1}^{i} W_{-1}^{j} ,
\end{equation}

\[\partial _{i} W_{-1}^{i} =0.\]
The system of equations \eqref{EQ17} and \eqref{EQ18} gives the secular terms

\begin{equation} \label{EQ16}
  -\nabla _{i} P_{-3} =D^{j} \varepsilon _{ijk} W_{-1}^{k} ,
\end{equation}
which corresponds to a geostrophic equilibrum equation.

In zero order $R^{0} $, we have the following system of equations:
\[ \partial _{t} v_{0}^{i} -\partial _{jj} v_{0}^{i} +\partial _{j} (W_{-1}^{i} v_{0}^{j} +v_{0}^{i} W_{-1}^{j} )+D^{j} \varepsilon _{ijk} v_{0}^{k} =\]
\begin{equation} \label{EQ17}
   =-(\partial _{i} P_{0} +\nabla _{i} P_{-2} )+F_{0}^{i} ,
\end{equation}

\[\partial _{i} v_{0}^{i} =0.\]
These equations give one secular equation:

\begin{equation} \label{EQ18}
   \nabla P_{-2} =0,\Rightarrow P_{-2} =Const.
\end{equation}
Let us consider the equations of the first approximation $R$:
\[{\partial _{t} v_{1}^{i} -\partial _{jj} v_{1}^{i} +D^{j} \varepsilon _{ijk} v_{1}^{k} +\partial _{j} (W_{-1}^{i} v_{1}^{j} +v_{1}^{i} W_{-1}^{j} +v_{0}^{i} v_{0}^{j} )=}\]

\begin{equation} \label{EQ19}
 =-\nabla _{j} (W_{-1}^{i} W_{-1}^{j} )-(\partial _{i} P_{1} +\nabla _{i} P_{-1} ),
\end{equation}

\begin{equation} \label{EQ20}
  \partial _{i} V_{1}^{i} +\nabla _{i} W_{-1}^{i} =0.
\end{equation}
From this system of equations there follows the secular equations:

\begin{equation} \label{EQ21}
   \nabla _{i} W_{-1}^{i} =0,
\end{equation}

\begin{equation} \label{EQ22}
   \nabla _{j} (W_{-1}^{i} W_{-1}^{j} )=-\nabla _{i} P_{-1} ,
\end{equation}
The secular equations \eqref{EQ27} and \eqref{EQ29} are satisfied by choosing the following geometry for the velocity field ( Beltrami field):

\begin{equation} \label{EQ23}
   \vec{W}=(W_{-1}^{x} (Z),W_{-1}^{y} (Z),0);
\end{equation}

\[\nabla P_{-1} =0,\Rightarrow P_{-1} =Const.\]
In the second order $R^{2}$, we obtain the equations
\[\partial _{t} v_{2}^{i} -\partial _{jj} v_{2}^{i} -2\partial _{j} \nabla _{j} v_{0}^{i} +\partial _{j} (W_{-1}^{i} v_{2}^{j} +v_{2}^{i} W_{-1}^{j} +v_{0}^{i} v_{1}^{j} +v_{1}^{i} v_{0}^{j} )\]
\begin{equation} \label{EQ24}
  +D^{j} \varepsilon _{ijk} v_{2}^{k} =-\nabla _{j} (W_{-1}^{i} v_{0}^{j} +v_{0}^{i} W_{-1}^{j} )-(\partial _{i} P_{2} +\nabla _{i} P_{0} ),
\end{equation}

\begin{equation} \label{EQ25}
    \partial _{i} v_{2} +\nabla _{i} v_{0} =0.
\end{equation}
It is easy to see that there are no secular terms in this order.

Let us come now to the most important order $R^{3}$. In this order we obtain the equations

\begin{equation} \label{EQ26}
\partial _{t} v_{3}^{i} +\partial _{T} W_{-1}^{i} -(\partial _{jj} v_{3}^{i} +2\partial _{j} \nabla _{j} v_{1}^{i} +\nabla _{jj} W_{-1}^{i} )+\nabla _{j} (W_{-1}^{i} v_{1}^{j} +v_{1}^{i} W_{-1}^{j} +v_{0}^{i} v_{0}^{j} )+
\end{equation}

\[+\partial _{j} (W_{-1}^{i} v_{3}^{j} +v_{3}^{i} W_{-1}^{j} +v_{0}^{i} v_{2}^{j} +v_{2}^{i} v_{0}^{j} +v_{1}^{i} v_{1}^{j} )+D^{j} \varepsilon _{ijk} v_{3}^{k} =-(\partial _{i} P_{3} +\nabla _{i} \overline{P}_{1} ),\]

\[\partial _{i} v_{3} +\nabla _{i} v_{1} =0.\]

From this we get the main secular equation:

\begin{equation} \label{EQ27}
   \partial _{T} W_{-1}^{i} -\Delta W_{-1}^{i} +\nabla _{k} (\overline{v_{0}^{k} v_{0}^{i} })=-\nabla _{i} \overline{P}_{1} ,
\end{equation}
There is also an equation to find the pressure $P_{-3}$:

\begin{equation} \label{EQ28}
-\nabla _{i} P_{-3} =D^{j} \varepsilon _{ijk} W_{-1}^{k} .
\end{equation}

\section{The velocity field in zero approximation  }
\label{i4}

It is clear that the most important is equation \eqref{EQ36}. In order to obtain these equations in closed form, we need to calculate the Reynolds stresses $\nabla _{k} (\overline{v_{0}^{k} v_{0}^{i} })$. First of all we have to calculate the fields of zero approximation $v_{0}^{k} $. From the asymptotic development in zero order we have

\begin{equation} \label{EQ29}
  \partial _{t} v_{0}^{i} -\partial _{jj} v_{0}^{i} +W_{-1}^{k} \partial _{k} v_{0}^{i} +D^{j} \varepsilon _{ijk} v_{0}^{k} =-\partial _{i} P_{0} +F_{0}^{i} ,
\end{equation}
Let us introduce the operator $\widehat{D}_{0}$:

\begin{equation} \label{EQ30}
  \widehat{D}_{0} \equiv \partial _{t} -\partial _{jj} +W^{k} \partial _{k} .
\end{equation}
Using $\widehat{D}_{0} $, we rewrite Equations \eqref{EQ29}:

\begin{equation} \label{EQ31}
   \widehat{D}_{0} v_{0}^{i} +D^{j} \varepsilon _{ijk} v_{0}^{k} =-\partial _{i} P_{0} +F_{0}^{i} ,
\end{equation}
Pressure  $P_0$ can be found from condition $div\vec{V}=0$.

\begin{equation} \label{EQ32}
  P_{0} =\frac{\left[\vec{D}\times \vec{\partial }\right]_{i} v_{0} ^{i} }{\partial ^{2} }
\end{equation}
Let us introduce  designations for operators $\widehat{D}_0$:

\begin{equation} \label{EQ33}
  \widehat{P}_{ij} =\partial _{j} \frac{\left[\vec{D}\times \vec{\partial }\right]_{i} }{\partial ^{2} }
\end{equation}
and for velocities: $v_{0} ^{x} =u_{0}$, $v_{0} ^{y} =v_{0}$, $v_{0} ^{z} =w_{0}$. Then excluding pressure from \eqref{EQ31} , we obtain the system of equations to find the velocity field of zero approximation~:
\[{\left(\widehat{D}_{0} +\widehat{P}_{xx} \right)u_{0} +\left(\widehat{P}_{yx} -D_{z} \right)v_{0} +\left(\widehat{P}_{zx} +D_{y} \right)w_{0} =F_{0} ^{x} ,}\]
\[{\left(\widehat{P}_{xy} +D_{z} \right)u_{0} +\left(\widehat{D}_{0} +\widehat{P}_{yy} \right)v_{0} +\left(\widehat{P}_{zy} -D_{x} \right)w_{0} =F_{0} ^{y} ,}\]

\begin{equation} \label{EQ34}
  {\left(\widehat{P}_{xz} -D_{y} \right)u_{0} +\left(\widehat{P}_{yz} +D_{x} \right)v_{0} +\left(\widehat{D}_{0} +\widehat{P}_{zz} \right)w_{0} =F_{0} ^{z} .}
\end{equation}
For simplicity, we choose the systeme of coordinates so that the axis $Z$ coincides with the direction of angular velocity of rotation $\vec{\Omega }$. Then $D_{x} =0$,  $D_{y} =0$,  $D_{z} =D$. In order to solve this system of equations we have to set the force in the explicit form. Let us choose now the external force in the rotating system of coordinates in the following  form:
\[F_{0} ^{z} =0,\quad \vec{F}_{0\bot } = f_{0} \left(\vec{i} \cos\phi _{2} +\vec{j} \cos\phi _{1} \right);\]
\[ \phi _{1} =\vec{k}_{1} \vec{x}-\omega _{0} t, \quad \phi _{2} =\vec{k}_{2} \vec{x}-\omega _{0} t, \]
\[{\vec{k}_{1} =k_{0} \left(1,0,1\right),\quad \vec{k}_{2} =k_{0} \left(0,1,1\right).}\]
It is obvious that divergence of this force us equal to zero. Thus, external force is given in plane $(x,y)$, orthogonal to rotation axis.

The solution for equations system \eqref{EQ34} can be found easily in accordance with Cramer's Rule:

\begin{equation} \label{EQ35}
  u_{0} =\frac{\Delta _{1} }{\Delta } ,v_{0} =\frac{\Delta _{2} }{\Delta } ,w_{0} =\frac{\Delta _{3} }{\Delta } .
\end{equation}

Here $\Delta $-is the determinant of the system \eqref{EQ34}:

\begin{equation} \label{EQ36}
  \Delta =\left|\begin{array}{ccc} {\widehat{D}_{0} +\widehat{P}_{xx} } & {\widehat{P}_{yx} -D} & {\widehat{P}_{zx} } \\ {\widehat{P}_{xy} +D} & {\widehat{D}_{0} +\widehat{P}_{yy} } & {\widehat{P}_{zy} } \\ {\widehat{P}_{xz} } & {\widehat{P}_{yz} } & {\widehat{D}_{0} +\widehat{P}_{zz} } \end{array}\right|,
\end{equation}

\begin{equation} \label{EQ37}
  \Delta _{1} =\left|\begin{array}{ccc} {F_{0} ^{x} } & {\widehat{P}_{yx} -D} & {\widehat{P}_{zx} } \\ {F_{0} ^{y} } & {\widehat{D}_{0} +\widehat{P}_{yy} } & {\widehat{P}_{zy} } \\ {0} & {\widehat{P}_{yz} } & {\widehat{D}_{0} +\widehat{P}_{zz} } \end{array}\right|,
\end{equation}

\begin{equation} \label{EQ38}
  \Delta _{2} =\left|\begin{array}{ccc} {\widehat{D}_{0} +\widehat{P}_{xx} } & {F_{0} ^{x} } & {\widehat{P}_{zx} } \\ {\widehat{P}_{xy} +D} & {F_{0} ^{y} } & {\widehat{P}_{zy} } \\ {\widehat{P}_{xz} } & {0} & {\widehat{D}_{0} +\widehat{P}_{zz} } \end{array}\right|,\;
\end{equation}

\begin{equation} \label{EQ39}
   \Delta _{3} =\left|\begin{array}{ccc} {\widehat{D}_{0} +\widehat{P}_{xx} } & {\widehat{P}_{yx} -D} & {F_{0} ^{x} } \\ {\widehat{P}_{xy} +D} & {\widehat{D}_{0} +\widehat{P}_{yy} } & {F_{0} ^{y} } \\ {\widehat{P}_{xz} } & {\widehat{P}_{yz} } & {0} \end{array}\right|.
\end{equation}
After writing down the determinants in the explicit form, we obtain:

\begin{equation} \label{EQ40}
  \begin{array}{l} {u_{0} =\frac{1}{\Delta } \left[\left(\widehat{D}_{0} +\widehat{P}_{yy} \right)\left(\widehat{D}_{0} +\widehat{P}_{zz} \right)-\left(\widehat{P}_{yz} \right)\left(\widehat{P}_{zy} \right)\right]F_{0} ^{x} +} \\ {+\frac{1}{\Delta } \left[\left(\widehat{P}_{zx} \right)\left(\widehat{P}_{yz} \right)-\left(\widehat{P}_{yx} -D\right)\left(\widehat{D}_{0} +\widehat{P}_{zz} \right)\right]F_{0} ^{y} ,} \end{array}
\end{equation}

\begin{equation} \label{EQ41}
  \begin{array}{l} {v_{0} =\frac{1}{\Delta } \left[\left(\widehat{P}_{xz} \right)\left(\widehat{P}_{zy} \right)-\left(\widehat{P}_{xy} +D\right)\left(\widehat{D}_{0} +\widehat{P}_{zz} \right)\right]F_{0} ^{x} +} \\ {+\frac{1}{\Delta } \left[\left(\widehat{D}_{0} +\widehat{P}_{xx} \right)\left(\widehat{D}_{0} +\widehat{P}_{zz} \right)-\left(\widehat{P}_{xz} \right)\left(\widehat{P}_{zx} \right)\right]F_{0} ^{y} ,} \end{array}
\end{equation}

\begin{equation} \label{EQ42}
  \begin{array}{l} {w_{0} =\frac{1}{\Delta } \left[\left(\widehat{P}_{xy} +D\right)\left(\widehat{P}_{yz} \right)-\left(\widehat{P}_{xz} \right)\left(\widehat{D}_{0} +\widehat{P}_{yy} \right)\right]F_{0} ^{x} +} \\ {+\frac{1}{\Delta } \left[\left(\widehat{P}_{xz} \right)\left(\widehat{P}_{yx} -D\right)-\left(\widehat{D}_{0} +\widehat{P}_{xx} \right)\left(\widehat{P}_{yz} \right)\right]F_{0} ^{y} ,} \end{array}
\end{equation}

\begin{equation} \label{EQ43}
  \begin{array}{l} {\Delta =\left(\widehat{D}_{0} +\widehat{P}_{xx} \right)\left[\left(\widehat{D}_{0} +\widehat{P}_{yy} \right)\left(\widehat{D}_{0} +\widehat{P}_{zz} \right)-\left(\widehat{P}_{yz} \right)\left(\widehat{P}_{zy} \right)\right]-} \\ {-\left(\widehat{P}_{yx} -D\right)\left[\left(\widehat{P}_{xy} +D\right)\left(\widehat{D}_{0} +\widehat{P}_{zz} \right)-\left(\widehat{P}_{xz} \right)\left(\widehat{P}_{zy} \right)\right]+} \\ {+\left(\widehat{P}_{zx} \right)\left[\left(\widehat{P}_{xy} +D\right)\left(\widehat{P}_{yz} \right)-\left(\widehat{D}_{0} +\widehat{P}_{yy} \right)\left(\widehat{P}_{xz} \right)\right].} \end{array}
\end{equation}
In order to calculate the expressions \eqref{EQ40}-\eqref{EQ43} we present the external force in complex form:

\begin{equation} \label{EQ44}
  F_{0} ^{x} =\frac{f_{0} }{2} \left(e^{i\phi _{2} } +e^{-i\phi _{2} } \right), \quad F_{0} ^{y} =\frac{f_{0} }{2} \left(e^{i\phi _{1} } +e^{-i\phi _{1} } \right).
\end{equation}
Then all operators in formulae   \eqref{EQ40}-\eqref{EQ42} act from the left on their eigenfunctions. In particular:

\begin{equation} \label{EQ45}
   \begin{array}{l} {\widehat{D}_{0} e^{i\phi _{2} } =e^{i\phi _{2} } \widehat{D}_{0} \left(\vec{k_{2} },  -\omega _{0} \right), \quad \widehat{D}_{0} e^{i\phi _{1} } =e^{i\phi _{1} } \widehat{D}_{0} \left(\vec{k_{1} },-\omega _{0} \right),} \\ {\Delta e^{i\phi _{2} } =e^{i\phi _{2} } \Delta \left(\vec{k}_{2} ,-\omega _{0} \right), \quad \Delta e^{i\phi _{1} } =e^{i\phi _{1} } \Delta \left(\vec{k}_{1} , -\omega _{0} \right).} \end{array}
\end{equation}
To simplify the formulae, let us choose $k_{0} =1$, $\omega _{0} =1$.

We will designate
\[ \widehat{D}_{0} \left(\vec{k}_{2} , -\omega _{0} \right)=2+i\left(w_{y} -1\right)=A_{y} ,\]
\begin{equation}\label{EQ46}
   \widehat{D}_{0} \left(\vec{k}_{1} , -\omega _{0} \right)=2+i\left(w_{x} -1\right)=A_{x} .
\end{equation}
Before do further calculations, we h ave to note that some components of tensors  $\widehat{P}_{ij} \left(\vec{k}_{1} \right)$  and  $\widehat{P}_{ij} \left(\vec{k}_{2} \right)$ vanish. Let us write the non-zero components only:
\[\widehat{P}_{yx} \left(\vec{k}_{1} \right)=\frac{1}{2} D, \quad \widehat{P}_{xz} \left(\vec{k}_{2} \right)=-\frac{1}{2} D, \quad \widehat{P}_{xy} \left(\vec{k}_{2} \right)=-\frac{1}{2} D,\]
\begin{equation} \label{EQ47}
  \widehat{P}_{yz} \left(\vec{k}_{1} \right)=\frac{1}{2} D.
\end{equation}
Taking into account the formulae \eqref{EQ45}-\eqref{EQ47}, we can find the determinant:

\begin{equation} \label{EQ48}
  \Delta \left(\vec{k}_{1} \right)=A_{x} ^{3} +\frac{1}{2} D^{2} A_{x},\quad  \Delta \left(\vec{k_{2} }\right)=A_{y} ^{3} +\frac{1}{2} D^{2} A_{y} .
\end{equation}
In a similar way we find velocity field of zero approximation:

\begin{equation} \label{EQ49}
  u_{0} =f_{0} \frac{A_{y} }{2} \frac{e^{i\phi _{2} } }{A^{2} _{y} +\frac{1}{2} D^{2} } +f_{0} \frac{D}{4} \frac{e^{i\phi _{1} } }{A^{2} _{x} +\frac{1}{2} D^{2} } +C.C.,
\end{equation}

\begin{equation} \label{EQ50}
  v_{0} =-f_{0} \frac{D}{4} \frac{e^{i\phi _{2} } }{A_{y} ^{2} +\frac{1}{2} D^{2} } +f_{0} \frac{A_{x} }{2} \frac{e^{i\phi _{1} } }{A_{x} ^{2} +\frac{1}{2} D^{2} } +C.C.,
\end{equation}

\begin{equation} \label{EQ51}
   w_{0} =f_{0} \frac{D}{4} \frac{e^{i\phi _{2} } }{A^{2} _{y} +\frac{1}{2} D^{2} } -f_{0} \frac{D}{4} \frac{e^{i\phi _{1} } }{A^{2} _{x} +\frac{1}{2} D^{2} } +C.C..
\end{equation}

\section{Reynolds stress and large scale instability}
\label{i5}

 To close the equations \eqref{EQ27} we have to calculate the Reynolds stresses $\overline{w_{0} u_{0}}$ and $\overline{w_{0} v_{0}}$ .

These terms are easily calculated with help of formulae \eqref{EQ49}-\eqref{EQ51}. As a result we obtain:
\[{\overline{w_{0} u_{0} }=\frac{f^{2} _{0} }{2} \frac{D}{\left|A^{2} _{y} +\frac{1}{2} D^{2} \right|^{2} } -\frac{f^{2} _{0} }{8} \frac{D^{2} }{\left|A^{2} _{x} +\frac{1}{2} D^{2} \right|^{2} } ,}\]
\begin{equation}\label{EQ52}
    {\overline{w_{0} v_{0} }=-\frac{f^{2} _{0} }{8} \frac{D^{2} }{\left|A^{2} _{y} +\frac{1}{2} D^{2} \right|^{2} } -\frac{f^{2} _{0} }{2} \frac{D}{\left|A^{2} _{x} +\frac{1}{2} D^{2} \right|^{2} } .}
\end{equation}
Now equations \eqref{EQ27} are closed and take form:
\[{\partial _{T} W_{x} -\Delta W_{x} +\frac{\partial }{\partial z} \overline{w_{0} u_{0} }=0,}\]

\begin{equation} \label{EQ53}
 {\partial _{T} W_{y} -\Delta W_{y} -\frac{\partial }{\partial z} \overline{w_{0} v_{0} }=0.}
\end{equation}
We calculate the modules and write  the Reynolds stresses \eqref{EQ52} in the explicit form:
\[{\; \; \; \; \overline{w_{0} u_{0} }=\frac{f^{2} _{0} }{2} \frac{D}{16\left(w_{y} -1\right)^{2} +\left[4+\frac{1}{2} D^{2} -\left(w_{y} -1\right)^{2} \right]^{2} } -}\]
\[{-\frac{f^{2} _{0} }{8} \frac{D^{2} }{16\left(w_{x} -1\right)^{2} +\left[4+\frac{1}{2} D^{2} -\left(w_{x} -1\right)^{2} \right]^{2} } , }\]
\[{\overline{w_{0} v_{0} }=-\frac{f^{2} _{0} }{8} \frac{D^{2} }{16\left(w_{y} -1\right)^{2} +\left[4+\frac{1}{2} D^{2} -\left(w_{y} -1\right)^{2} \right]^{2} } -}\]
\begin{equation} \label{EQ54} {-\frac{f^{2} _{0} }{2} \frac{D}{16\left(w_{x} -1\right)^{2} +\left[4+\frac{1}{2} D^{2} -\left(w_{x} -1\right)^{2} \right]^{2} }  . } \end{equation}
With small  $W_{x}$, $W_{y}$ Reynolds stresses \eqref{EQ52} can be expanded in a series in the small parameters$W_{x}$, $W_{y} $. Taking into account the formula:

\[\frac{1}{\left|A^{2} _{x,y} +\frac{1}{2} D^{2} \right|^{2} } =Const -\frac{32\left(D^{2} -10\right)}{\left[\left(D^{2} +6\right)^{2} +64\right]^{2} } w_{x,y} +\cdots \]
we obtain the linearized equations \eqref{EQ53}:
\[{\; \frac{\partial }{\partial T} W_{x} -\frac{\partial ^{2} }{\partial z^{2} } W_{x} -\frac{\alpha f^{2} _{0} D}{2} \frac{\partial }{\partial z} W_{y} +\frac{\alpha f^{2} _{0} D^{2} }{8} \frac{\partial }{\partial z} W_{x} =0,} \]
\begin{equation}\label{EQ55}
    {\frac{\partial }{\partial T} W_{y} -\frac{\partial ^{2} }{\partial z^{2} } W_{y} +\frac{\alpha f^{2} _{0} D^{2} }{8} \frac{\partial }{\partial z} W_{y} +\frac{\alpha f^{2} _{0} D}{2} \frac{\partial }{\partial z} W_{x} =0.}
\end{equation}
\[\alpha =\frac{32\left(10-D^{2} \right)}{\left[\left(D^{2} +6\right)^{2} +64\right]^{2} } .\]
We will search for the solution of linear system \eqref{EQ55} in the form:

\begin{equation} \label{EQ56}
  W_{x} ,W_{y} \sim \exp \left(\gamma T+ikZ\right).
\end{equation}
We substitute \eqref{EQ56} in equation \eqref{EQ55} and obtain the dispersion equation:

\begin{equation} \label{EQ57}
  \gamma =-ik\frac{\alpha f^{2} _{0} D^{2} }{8} \pm k\frac{\alpha f^{2} _{0} D}{2} -k^{2} .
\end{equation}
The dispersion equation \eqref{EQ57} shows that equation system \eqref{EQ55} has instable oscillatory solutions with  oscillatory frequency   $\omega =k\frac{\alpha f^{2} _{0} D^{2} }{8}$ and instability growth rate  $\gamma =k\frac{\alpha f^{2} _{0} D}{2} -k^{2}$. The instability is large scale because the instable term dominates over dissipation  on large scales:  $\frac{\alpha f^{2} _{0} D}{2} > k$. The maximum growth rate of instability is equal to   $\gamma _{\max } = \frac{\alpha ^{2} f^{4} _{0} D^{2} }{16}$,   and is achieved on the wave vector  $k_{\max } =\frac{\alpha f^{2} _{0} D}{4}$. As a result of the development of instability  the large scale helical circular polarized vortices of Beltrami type are generated in the system.

\section{Saturation of instability and nonlinear vortex structures}
\label{i6}

\begin{figure}
  \centering
  \includegraphics[width=7 cm]{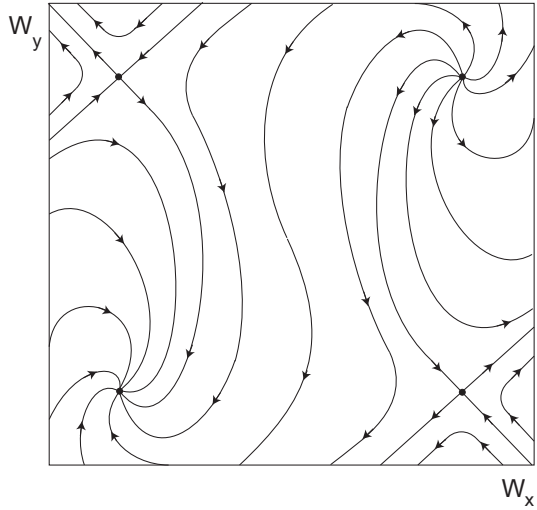}\\
  \caption{Phase portrait of the dynamical system \eqref{EQ58}, with $D=1$, $C_1 =-0.03$, $C_2 =0.03$. One can see two hyperbolic singular points and stable and instable knots.}\label{f1}
\end{figure}

It is clear that with increasing of amplitude nonlinear terms decrease and instability becomes saturated. Consequently stationary nonlinear vortex structures are formed.  To find these structures let us choose for equations \eqref{EQ54} $\frac{\partial }{\partial T} =0$ and integrate equations one time over $Z$. We obtain the system of equations:
\[{\; \frac{d}{dZ} \; W_{x} =\overline{w_{0} u_{0} }+C_{1} ,}\]
\begin{equation} \label{EQ58}
  {\frac{d}{dZ} \; W_{y} =\overline{w_{0} v_{0} }+C_{2} .\; }
\end{equation}
From equations \eqref{EQ58} follows:

\begin{equation} \label{EQ59}
   \frac{dw_{x} }{dw_{y} } \; =\frac{\overline{w_{0} u_{0} }+C_{1} }{\overline{w_{0} v_{0} }+C_{2} } ,
\end{equation}
After integrating the system of equations \eqref{EQ59} we obtain:
\begin{figure}
  \centering
  \includegraphics[width=7 cm]{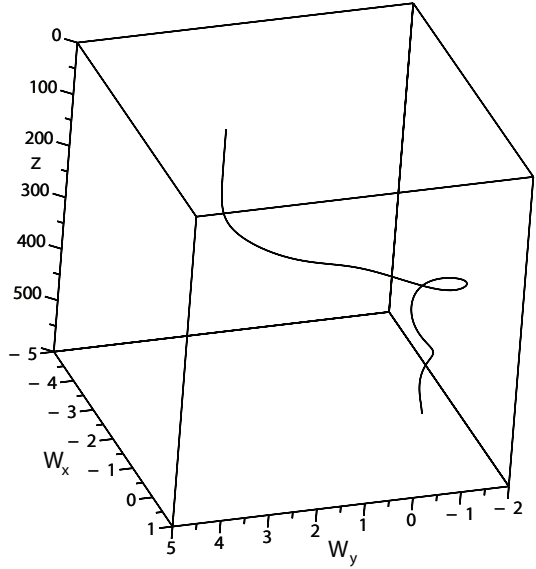}\\
  \caption{The kink which connects the hyperbolic point with stable knot with $D=1$, $C_1=0.04$, $C_2=0.04$. When approaching the stable knot one can see rotations of velocity field.}\label{f2}
\end{figure}

\begin{equation} \label{EQ60}
   \int \overline{w_{0} v_{0} }dw_{x}  +C_{2} w_{x} =\int \overline{w_{0} u_{0} }dw_{y} +C_{1} w_{y}  .
\end{equation}
Integrals in expression \eqref{EQ60} are calculated in elementary functions (see \cite{s17}), which give the expression for first integral of motion $J$ of equations \eqref{EQ59}:
\[{J=\frac{D^{2} }{8} \frac{w_{x} }{\left[4+\frac{1}{2} D^{2} -\left(w_{y} -1\right)^{2} \right]^{2} +16\left(w_{y} -1\right)^{2} } +}\]
\[{+\frac{D}{2^{5/2} \left(8+D^{2} \right)} \ln \frac{\left(w_{x} -1\right)^{2} +\left(w_{x} -1\right)D\sqrt{2} +4+\frac{1}{2} D^{2} }{\left(w_{x} -1\right)^{2} -\left(w_{x} -1\right)D\sqrt{2} +4+\frac{1}{2} D^{2} } +}\]
\[{+\frac{D}{8\left(8+D^{2} \right)} \arctan \frac{\left(w_{x} -1\right)^{2} -4-\frac{1}{2} D^{2} }{4\left(w_{x} -1\right)} -}\]
\[{-\frac{D^{2} }{8} \frac{w_{y} }{\left[4+\frac{1}{2} D^{2} -\left(w_{x} -1\right)^{2} \right]^{2} +16\left(w_{x} -1\right)^{2} } +}\]
\[{+\frac{D}{2^{5/2} \left(8+D^{2} \right)} \ln \frac{\left(w_{y} -1\right)^{2} +\left(w_{y} -1\right)D\sqrt{2} +4+\frac{1}{2} D^{2} }{\left(w_{y} -1\right)^{2} -\left(w_{y} -1\right)D\sqrt{2} +4+\frac{1}{2} D^{2} } +}\]
\[{+\frac{D}{8\left(8+D^{2} \right)} \arctan \frac{\left(w_{y} -1\right)^{2} -4-\frac{1}{2} D^{2} }{4\left(w_{y} -1\right)} +C_{1} w_{y} +C_{2} w_{x} .}\]
Equations \eqref{EQ58} can be easily calculated numerically using standard tools. In particular, this allows to construct phase portrait of the dynamical system \eqref{EQ58} ( fig.\ref{f1}) and to get the most interesting solutions which link singular points on phase plane.  See for example fig.\ref{f2}, where the hyperbolic singular point is connected with the stable knot and fig.3, where the solution connects instable and stable focuses. All these solutions correspond to the large scale localized vortex structures of kink type with rotation, generated by the instability which has been found in this work.
\begin{figure}
  \centering
  \includegraphics[width=7 cm]{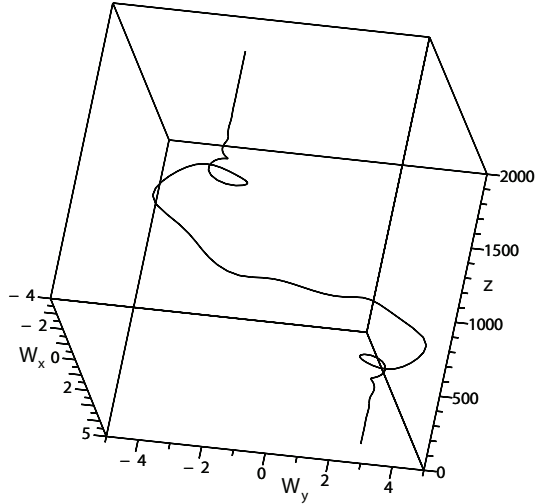}\\
  \caption{The kink which connects the instable and stable focuses with $D=1$, $C_1=0.04$, $C_2=0.04$. One can see the internal helical structure of the kink.}\label{f3}
\end{figure}

\section{Conclusions and discussion of the results}
\label{i7}

In this work we find new large scale instability in rotating fluid.  It is supposed that the small scale vortex external force in rotating coordinates system acts on fluid which maintains the small velocity field fluctuations (small scale turbulence with small Reynolds number $R$, $R\ll 1$ ).  For the real applications this Reynolds number should be calculated with help of the turbulent viscosity. The asymptotic development of motion equations by small Reynolds number   allows obtaining motion equations for the large scale. These equations are of the hydrodynamic $\alpha $-effect type, in which velocity components $W_{x}$, $W_{y} $ are connected by the positive feedback. This may result in the appearance of the large scale vortex instability.  The large scale vortices of Beltrami type are formed due to this instability in rotating fluid with small scale exterior force. With further increase of amplitude the instability stabilizes and passes to a stationary mode. In this mode the nonlinear stationary vortex structures form. Different vortex kinks belong to the most interesting structures.  These kinks link stationary points of dynamical system \eqref{EQ58}. The kink which links hyperbolic point with stable knot has rotations around the stable knot as shown on fig.\ref{f2}. In the kink which links instable and stable focuses, vector field turns around two singular points, see fig.\ref{f3}.

Let us note that unlike previous works about hydrodynamic $\alpha $- effect in rotating fluid, the use of the asymptotic development allows to construct naturally the nonlinear theory and to study the stationary nonlinear vortex kinks.

\end{document}